# Heterogeneity-aware Fault Tolerance using a Self-Organizing Runtime System

Mario Kicherer and Wolfgang Karl
Chair for Computer Architecture and Parallel Processing
Karlsruhe Institute of Technology
Email: {kicherer,karl}@kit.edu

*Abstract*—Due to the diversity and implicit redundancy in terms of processing units and compute kernels, off-the-shelf heterogeneous systems offer the opportunity to detect and tolerate faults during task execution in hardware as well as in software. To automatically leverage this diversity, we introduce an extension of an online-learning runtime system that combines the benefits of the existing performance-oriented task mapping with task duplication, a diversity-oriented mapping strategy and heterogeneity-aware majority voter. This extension uses a new metric to dynamically rate the remaining benefit of unreliable processing units and a memory management mechanism for automatic data transfers and checkpointing in the host and device memories.

## I. INTRODUCTION

Accelerators like GPUs promise a significant performance improvement for certain problems compared to the calculation on a general-purpose CPU. However, they also increase application complexity through individual programming models and software stacks for the compute kernels, e.g., dedicated compiler and runtime libraries, and expensive data transfers to and from device memory. Due to the resulting increase in source code complexity and the additional layers in the software stack required for building and successfully executing the application, the probability of faults caused by software bugs or incompatibility grows. To make things worse, the susceptibility of the hardware to faults is expected to increase as well: due to shrinking feature sizes, aging effects and charged particles hitting conductor paths could become a considerable threat for calculations [15]. Consequently, application execution will depend on a growing number of processing units with decreasing reliability and on growing middleware layers that introduce further complexity on their own [9].

However, if properly incorporated, heterogeneity also offers opportunities to increase the reliability of application execution. On homogeneous systems, redundancy-based approaches to detect faults in hardware are well-known but detecting a fault caused by a bug in the source code using redundant execution is difficult as a calculation returns the same erroneous result on any processing unit. In heterogeneous systems, however, an application usually contains individual compute kernels for the different types of processing units, e.g., an OpenMP kernel for multiprocessors and a CUDA kernel for NVIDIA GPUs. Hence, by executing a calculation redundantly with the already existing compute kernels on different types of processing units, a bug in the source code as well as faults in the hardware can be detected.

While redundant execution and N-version programming are already known concepts, this work proposes to exploit the existing diversity for fault tolerance and uncovers challenges that have to be solved in order to establish efficient fault-tolerant compute kernel execution in heterogeneous systems with only marginal help of application developers. Similar to related projects, we use an online-learning runtime system for performance-oriented task mapping as basis for this work. We show how the mapping mechanism has to be extended in order to not only consider the performance of a processing unit but also its susceptibility to faults, how to efficiently manage the data in the dedicated device memories to minimize costly data transfers and how to extend the majority voter to consider the specialties of heterogeneous systems.

The remainder of this paper is structured as follows: we will first give an overview of related work and state of the art in Section 2. Afterwards, we describe the major techniques of our contribution. Performance and feasibility of this approach are evaluated and compared in Section 4. Finally, Section 5 concludes the paper giving further outlook.

## II. RELATED WORK

Dependability is a wide research topic with a long history. In this paper, we focus on work related to fault detection and tolerance in modern systems. In the following, we start with the related work focusing on reliability for CPU computations.

Many research projects propose fine-grained on-chip redundancy to decrease the costs for rollbacks and to benefit from underutilized resources. Targeting general-purpose CPUs, several projects utilize the features of modern processors, e.g., multiple cores and superscalar out-of-order pipelines [7], [12], [14].

Vera et al. [17] also propose a fine-grained redundancy approach for CPUs. They argue that only 20% of the instructions of a modern architecture are responsible for more than 60% of the total vulnerability. They introduce so-called selective replication of only certain instructions and achieve a considerable fault coverage while introducing only minor overhead. A similar approach based on VLIW architectures is introduced by Lee et al. [10] that exploits empty slots for dynamic duplication.

As a software-based solution, Rebaudengo et al. present a source-to-source compiler creating redundancy on the source-code level [13]. Their efforts aim to detect transient faults causing data and program-flow corruption.

Besides reducing the overhead of redundant execution, other approaches try to avoid redundancy at all by detecting faults by other light-weight indicators, such as symptoms like anomalous application behavior detected by segmentation faults or an unusual rate of branch mispredicts or cache misses [6]. Such detection mechanisms save time, but come at the price of mispredictions or lower fault coverage.

Besides symptom-based fault detection, arithmetic codes can be used for validation [18]. Here, input values for calculations are modified in a way that the results can be validated using a checksum-like mechanism.

In heterogeneous systems, important tasks of the application are migrated to accelerators and only protecting the computations on the CPU is not sufficient. Therefore, other projects present their efforts to increase reliability of heterogeneous computing.

Takizawa et al. introduce CheCUDA that enables a checkpoint and restart mechanism for CUDA kernels [16]. In combination with a tool for CPU-bound application checkpointing, applications with CUDA kernels can be restarted after a fault or even be migrated to another host.

For redundancy-based fault detection on GPUs, Dimitrov et al. [4] introduce and evaluate three possible methods to efficiently execute kernel code multiple times: simple duplication of kernel computations, interleaved kernel instructions, and exploiting unused thread-level parallelism. Like the CPU mechanisms, their efforts concentrate on a single type of accelerator. However, our approach can be used with arbitrary types of accelerators.

Another work targeting GPUs is from Fang et al. [5]. They introduce their debugger-based fault injector GPU-Qin that enables injections on instruction level. In their evaluation, they show that there are different classes of applications that exhibit a similar low or high susceptibility for corrupted results or abortion of execution.

Generic approaches not targeting a certain type of processing unit are proposed as well. Zhang et al. presented a mechanism for efficiently hiding faulty cores in a manycore processor [19]. Their solution maintains a sane view of a logical topology that does not only hide faulty cores but also improves the alignment of the cores for minimal communication costs.

Another approach for increasing reliability is reducing hardware-fault susceptibility: Mitra [11] proposes hardware-level techniques for reducing the susceptibility of circuits and predicting faults induced by infant mortality or aging. Also, he introduces special test patterns for online self-tests.

In contrast to the described approaches, our design does not depend on specific programming models or hardware. Existing kernels can be reused with only simple source code modifications and neither special hardware mechanisms nor special compilers are required. However, additional techniques like the ones described in this section are required to protect the execution of the other parts of the application, the runtime system itself and the operating system on the CPU.

Regarding performance portability in heterogeneous computing, several other projects exist that provide automatic task mapping [1], [2]. However, none of them considers dependability and the challenges of reliable task execution.

III. PRELIMINARY WORK

In preliminary work, we introduced an online-learning runtime system with a mechanism for performance-oriented task mapping in heterogeneous systems [8], [9]. This mechanism measures the time consumption of kernel executions and stores these values per kernel, problem size and processing unit in a system-wide database. This data can be used concurrently by different applications on the system to search for performance data gathered by other processes, thus avoiding redundant learning. Before execution of another task, the mechanism compares the runtimes of the kernels on the respective processing units and then chooses the fastest combination of kernel and processing unit for execution.

IV. TASK-LEVEL REDUNDANCY IN HETEROGENEOUS SYSTEMS

In the following two sections, we describe the challenges and solutions that are important to consider for efficient and fault-tolerant task execution on heterogeneous systems. Afterwards, we first show how to leverage the introduced solutions for light-weight fault tolerance with minor performance overhead at the expense of limited coverage of faults. Finally, we describe the design of our diversity-oriented task mapping that is able to detect faults in software and hardware and we give a short example how our approach is included in the source code.

*A. Memory management in heterogeneous systems*

A critical point in heterogeneous systems is efficient memory management as accelerators usually possess an own dedicated memory that implies expensive data transfers to and from host memory. Besides, these transfers usually have to be triggered manually by a developer. If multiple kernels on different processing units are executed one after another, the application developer is responsible for identifying the location of the most recent data and initiating transfers as necessary. Therefore, manual memory management complicates development and increase code complexity.

To simplify this, the runtime system contains an own memory management unit that simplifies development and automatically initiates data transfers only if necessary. After allocating and initializing a memory area, the developer can register the area at the runtime system and afterwards, a kernel can request the data and the runtime system will automatically transfer the data to the respective device memory of the processing unit or only return the new address, if the data is already present in this memory.

As memory can be used differently, e.g., as input or output, the kernels can state the type of access they request. If a kernel executing on an accelerator requests write access, the data in the original memory area in host memory becomes invalid. To keep track of these changes, the runtime system maintains a list of so-called siblings of an area in each memory and tags the areas with a version number that is increased by one on write access. Therefore, if a kernel requests certain data, the runtime system iterates through the list of siblings in order to find the sibling with the highest version. If this sibling is already in the corresponding memory, the pointer to this area is returned, if not, a new sibling is allocated and the data is transferred from the sibling with the highest version before returning the pointer to the kernel.

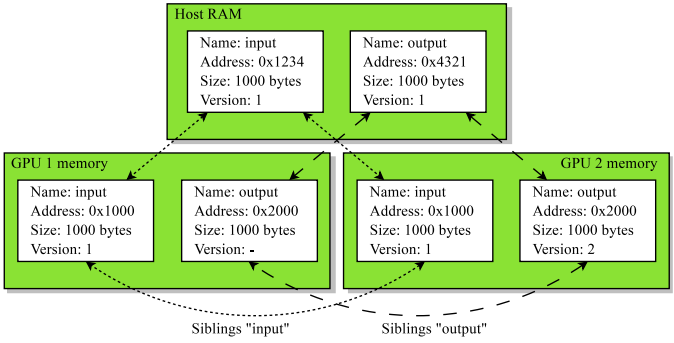

Fig. 1. State of memory areas after a faulty calculation on GPU 1

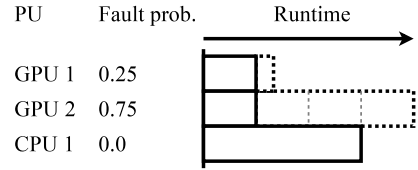

Fig. 2. Visual example for calculating the fault-aware runtime

Automatic memory management gets even more valuable for fault tolerant execution. If a fault occurs during kernel execution, a memory area can be in an undefined state and reusing this area for a second run could lead to corrupted results. For such cases, the memory management of the runtime system slightly differs. If there is an up-to-date sibling in the desired memory and another one in the same or another memory, the sibling in the desired memory is used like in the normal case. In case there is only one sibling with the highest version, a new sibling is created and returned to the kernel. If a fault is detected during execution of this kernel, the area can simply be marked as invalid and the runtime system can create a new copy of the other sibling. An example of such a case is illustrated in Figure 1. Initially, the developer registered the two memory areas "input" and "output". Then, the runtime system decides to start a kernel on GPU 1 and creates new siblings in the corresponding memory. During execution a fault occurs and "output" is invalidated. As we still have a up-to-date version of "output" in host RAM, the runtime system restarts execution on GPU 2. In this case, the execution succeeds and the runtime system increases the version of "output" in GPU 2 memory by one. If a kernel on another processing unit will afterwards request read access on "output", the data will be transferred from GPU 2 memory, as it has the highest version among all siblings.

### B. Fault-aware runtime estimation

Another important task in a heterogeneous system is the choice of the processing unit that returns the result of a calculation as fast as possible. However, the processing units in a heterogeneous system can be very versatile regarding performance but also regarding reliability due to different manufacturing processes, architectures and duty cycles.

Additionally, the dependability of a unit may change suddenly during application runtime, e.g., due to overheating caused by a failing fan. Therefore, statically choosing a certain type of processing unit, e.g., at compile time, is not advisable.

Instead of statically choosing a unit, the runtime system collects data about the processing units at runtime and uses this information to predict the fastest reliable unit for further runs. It stores the past execution times in relation to problem size and the past fault rate in its database that is shared with other processes. By comparing past execution times, the runtime system can simply find the fastest processing unit. However, if the reliability becomes a crucial factor, the important question for application performance is, how to trade off shorter execution times with the expected probability of a fault during a compute kernel execution – that would lead to an expensive restart of the kernel execution. For example, abandoning an accelerator that is more than two times faster as any other unit is not beneficial in case it once caused a fault. E.g., in case a transient fault occurred during the first run, the calculation can be repeated and the result is still returned sooner to the application than with any other processing unit. Therefore, the probability of faults during kernel execution must be combined with the expected performance benefits to determine if the processing unit is still beneficial in comparison to other units. On average, the accelerator in the example is still beneficial as long as its fault probability is below 50%.

Therefore, we propose the use of a simple metric called fault-aware runtime estimation that is calculated from the fault-free execution time and the observed fault rate of the processing unit. Specifically, the fault-aware runtime represents the fault-free kernel execution runtime plus the average time required until the processing unit calculates a correct result. We define the fault-aware runtime $F_i$ of the processing unit $i$ with its fault probability $p_i \in [0, 1)$, the number of past valid runs $v_i$, total number of runs $t_i$ and fault-free runtime $R_i$ as:

$$p_i = \frac{v_i}{t_i}$$

$$F_i = R_i * \frac{1}{1 - p_i}$$

A fault probability of 1 is handled like an infinite fault-aware runtime and the corresponding processing unit is only used in predefined check intervals to determine if it is still malfunctioning. We provide a visual example in Figure 2, where the solid boxes represent the fault-free runtime and the dashed boxes represent the fault-aware runtime for the given fault probability. In this example, GPU 2 is one of the fastest units but it is considered as the worst possible choice for task execution by the runtime system due to its fault probability of 0.75 which equals a fault-aware runtime that is four times higher as the fault-free runtime.

### C. Light-weight fault tolerance

As a trade-off between performance and the time-consuming redundancy-based methods described in the next section, the runtime system also provides light-weight fault detection and tolerance mechanisms that have only minor impact on the application runtime but also only limited fault coverage. Instead of choosing the processing unit with the lowest runtime, it chooses the processing unit with the lowest fault-aware runtime for execution in order to avoid unprofitable processing units. Before starting the execution, the runtime system ensures that a backup of the input data with the highest version exists and uses a separate thread to execute the kernel. With the separate thread and the backup copy, the runtime system has a checkpoint and it is able to rollback into this state if a fault occurs.

For example, if a direct fault like a segmentation fault occurs, the runtime system intercepts the normal fault handling – that would lead to an abortion of the whole application – and instead only stops the corresponding thread. In case of other direct faults, e.g., an error code returned by a function of the device's API, the data can be in an unknown state. Therefore, the runtime system invalidates the corresponding data in device memory and resets the thread's device context, if necessary. Afterwards, it starts the kernel execution again with the next best combination of kernel and processing unit.

Another problem that might occur during execution are non-responsive processing units or processing units that are trapped in an endless loop. Under normal conditions, detecting such cases is difficult as it is unknown, how much time a certain combination of kernel and processing unit takes on a specific system. However, as the runtime system maintains individual profiles with past execution times, it is able to estimate the runtime and to set an approximate timeout as product of runtime and an user-defined factor. Hence, in case of a non-responsive or non-stopping unit, the runtime system aborts the thread after a timeout and restarts execution with another combination of kernel and processing unit. Choosing an inappropriate timeout doesn't affect correctness and has only an impact on performance, as choosing the timeout too high or low only varies the number of restarts. We do not preset a timeout factor as a good value depends on the specific system, e.g., the possible amount of concurrent processes.

### D. Detecting and tolerating corrupted results

To detect corrupted results, the runtime system follows the dual-modular redundancy (DMR) concept. To detect faults in the hardware, it uses task duplication to spawn two redundant tasks that are mapped on the two combinations of kernels and processing units with the lowest fault-aware runtime. Afterwards, it starts the heterogeneity-aware majority voter to compare the results and determine the presumable correct result.

From the runtime system's perspective, this voter is again a regular task that can be executed with different compute kernels on different devices – if desired, it can be even executed redundantly itself. Similar to regular tasks, the runtime system can either choose the combination of kernel and processing unit with the lowest fault-aware runtime again or one of the remaining combinations that were not previously used for the actual task in order to avoid that calculation and comparison run on the same device.

While this mode can still be conservative, given there are enough unused resources, it is susceptible to faults in the code of the compute kernel as both processing units can use the same kernel and thus create the same erroneous result. Therefore, the runtime system also offers a so-called heterogeneous dual-modular redundancy (HetDMR) mode that enforces execution with different compute kernels on different processing units. Assuming that the kernels are sufficiently different, e.g., because they are created by different hardware experts, this mode enables the detection of faults in hardware as well as software.

The actual comparison of the results is done by bit-wise comparison of values in most cases, e.g., if the values are common integer types. However, if the values have a special type such as floating-point numbers, also special mechanisms are necessary for heterogeneous systems: as it is possible that the order of instructions differ between the different types of processing units and they may also use different rounding modes, the results may diverge although they are correct. For such a case, the programmer may define a maximum that determines how much two floating-point numbers are allowed to differ while they are still considered equal by the voter in the runtime system. During our experiments with OpenMP and CUDA implementations, the float values in the results differed by $0.001\%$ to $0.1\%$ depending on the application. Therefore, we set a delta of $0.1\%$ as acceptable for the calculations with single precision.

### E. Source code example

To give an example how our approach can be included in C programs, we provide the shortened source code of an application to increase an array of floats in Listing 1. Using a simple macro in Line 1, we define and initialize the function pointer `inc` with no return value (`void`) and three arguments of type `float*` and `int`. This pointer abstracts the actual task implementations for an application and it is called later instead of the actual `inc_CPU` or `inc_GPU` kernel function. As floating-point values require a special compare method, the corresponding memory and their types are registered using `dls_register_mem()` in Line 7 and 8. Finally in Line 10, the `inc` function pointer can be called like a usual function. Both kernels start with calls of `dls_request()`. Through this call, the data is automatically transferred into the memory of the processing unit and duplicated, if necessary, and the host address in the given variable is replaced with the new device address.

## V. EVALUATION

For evaluation, we used selected applications from the Rodinia benchmark suite [3] and the Nvidia CUDA SDK. For every Rodinia benchmark a separate OpenMP and CUDA application exists. To enable a dynamic switch between the two versions at runtime, we had to integrate both parts into one application. We did not consider benchmarks for the evaluation that are not suitable for the DMR concept, e.g., benchmarks where the OpenMP and the CUDA versions calculate different results, e.g., due to different data structures, or the versions make use of static variables which exclude a parallel execution of the functions.

For evaluation, we used a dual AMD Opteron 2378 machine equipped with 8 CPU cores, an Nvidia GeForce GTX 275 and a GeForce GTX 560Ti GPU running with Ubuntu

Listing 1. Code example for increasing an array

```
DLS_DECDEF(inc, void, float*, float*, int);              1

int main(int argc, char *argv[]) {                       3
  /* ... */                                              4

  size = sizeof(float)*count;                            6
  dls_register_data(input, size, DLS_VT_FLOAT, "r");     7
  dls_register_data(output, size, DLS_VT_FLOAT, "w");    8

  inc(input, output, count);                            10
}                                                       11

void inc_CPU(float * input, float * output, int c) {    13
  dls_request(&input, "r");                             14
  dls_request(&output, "w");                            15

  /* increase array on CPU */                           17
}                                                       18

void inc_GPU(float * input, float * output, int c) {    20
  dls_request(&input, "r");                             21
  dls_request(&output, "w");                            22

  /* increase array on GPU */                           24
}                                                       25
```

Linux 12.04. In order to decrease runtime variation due to competing tasks on the system, we limited the number of OpenMP threads to 7. Therefore, competing tasks like maintenance routines can execute on the eighth CPU core without increasing fluctuations of our measurements.

*A. Heterogeneity-aware Majority Voter*

Depending on the size of the data, comparing the results and determining the presumable correct result can have a significant impact on performance, especially if the results lie in different memories. Instead of statically executing the voter on the CPU, the runtime system contains kernels for different processing units and uses its mapping mechanism to determine the fastest reliable processing unit.

In Figure 3, we show the time consumption of the different kernels as function of the data size. As we can see, considering the data size is crucial for choosing the fastest kernel as the execution times can differ in one order of magnitude - without even considering the overhead for memory transfers. For small sizes, the single-threaded voter takes the least amount of time up until about 10 kB. Afterwards, the OpenMP kernel is the fastest for a small range of sizes until the massive-parallel GPU becomes the preferred choice for data bigger than 100 kB.

*B. Comparison of the different strategies*

In this experiment, we show the average time that is required until the runtime system returns a result to the application. First, we evaluated the raw execution time of the kernels on the different processing units and now show the results in Figure 4. Then, we measured the time with our runtime system and performance-oriented mapping (Perf), with performance-oriented mapping and checkpointing (Perf+CP) for light-weight fault tolerance, normal Dual-Modular Redundancy (DMR) and heterogeneous Dual-Modular Redundancy (HetDMR). As we can see in Figure 5, the checkpointing adds a small additional overhead for creating redundant copies of the output data and the DMR strategy increases the runtime by factor 1x to 3x. Except for the pathfinder benchmark, the HetDMR strategy enlarges the overhead significantly as the OpenMP kernel is considerably slower than the CUDA kernels. DMR and HetDMR are about the same for the pathfinder benchmark, as the OpenMP kernel is the fastest here and thus DMR and HetDMR choose the same processing units (the CPU and one GPU) for execution.

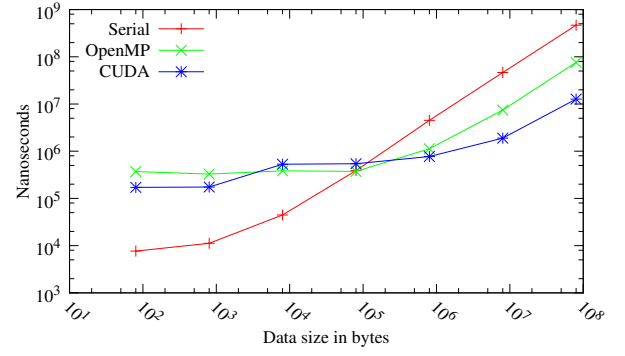

Fig. 3. Time consumption for comparison of two results with different size

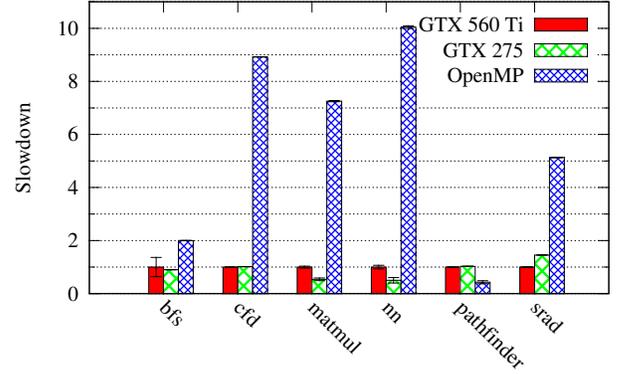

Fig. 4. Comparison of average kernel execution time on different processing units

*C. Benefit of the fault-aware runtime estimation*

In a further experiment, we twice measured the average time it takes our runtime system to return a valid result to the application with varying probability for a fault during execution of the pathfinder OpenMP kernel. One time, we used performance-oriented task mapping and checkpointing (Perf+CP) that chooses the kernel and processing unit with the lowest execution time. For the other part, we used our fault-aware runtime estimation to select the best task mapping. To simulate faults, we injected code at the end of the kernel that causes a segmentation fault under given conditions. As the OpenMP kernel is roughly 3x faster than the best GPU kernel, we expect our metric to accept the slower GPU kernel if the fault probability of the OpenMP mapping rises above $p = 1 - \frac{1}{3} = 66\%$. In Figure 6, we show that, as expected, the reliability-aware approach reduces the required time beginning with a fault probability of 70%. Compared to avoiding the CPU as soon as it caused a fault, our approach is, e.g., for 10% fault probability, more than 2 times faster.

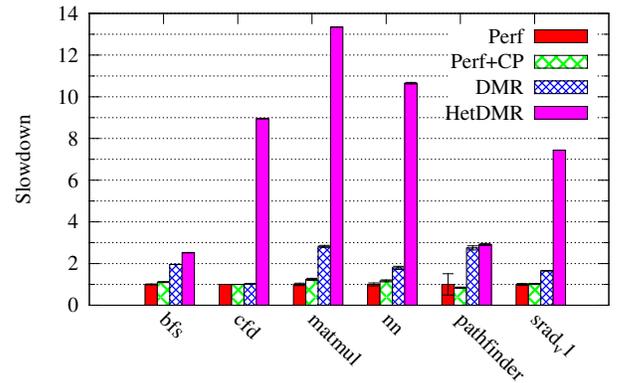

Fig. 5. Overhead of the three modes compared to performance-oriented task mapping

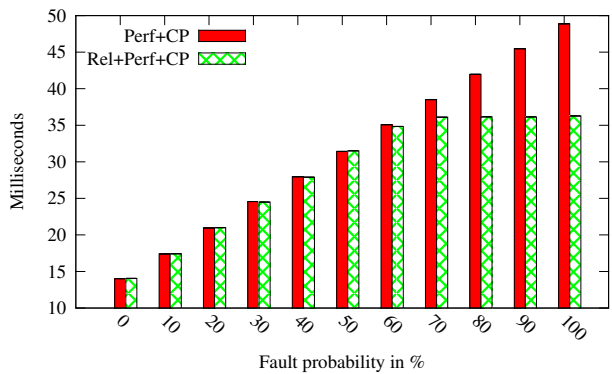

Fig. 6. Runtime with performance-oriented and fault-aware task mapping

## VI. Conclusion

In this work, we showed which challenges have to be solved in order to exploit design diversity and redundancy with only marginal additional efforts for developers. We extended an online-learning runtime system by mechanisms to automatically detect and tolerate different types of faults occurring during compute kernel execution. With only minor help of the developer, the runtime system preserves the view of fault-free compute kernel execution for the application although a compute kernel or processing unit might suffer from faults. We also introduced a new metric called fault-aware runtime estimation that is used to rate the remaining benefit of unreliable processing units and a memory management mechanism that automates data transfers and checkpointing. As energy efficiency is another important topic, we plan to extend this work further to also consider the energy consumption besides the runtime in future work.